\documentstyle[12pt]{article}

\begin{document}

\baselineskip=7mm

\begin{center}
{Clustering aspects and the shell model}

{A. Arima}

          The House of Councilors, 2-1-1 Nagatacho, Chiyodaku, 

               Tokyo 100-8962   Japan
 
\end{center}                

Abstract : ~
In this talk I shall discuss the clustering aspect and the shell model.
I shall first discuss the $\alpha$-cluster aspects based 
on the shell model calculations. Then I shall discuss the spin
zero ground state dominance in the presence of random interactions
and a new type of
cluster structure for fermions in a single-$j$ shell in the
presence of only pairing interaction with the largest multiplicity. 

\newpage

\section{INTRODUCTION}

As is well known, the clustering phenomenon
is universal in  nature. The $\alpha$ clustering picture 
in light  nuclei has been studied for many years. 
In this talk I shall first look back
at the history of the $\alpha$ clustering picture
in nuclear physics from the view point 
of the shell model \cite{Arimax}.  Next I shall introduce our efforts
towards understanding the origin of the
spin 0 ground state (0 g.s.)  dominance 
discovered by Johnson, Bertsch and Dean in 1998 using 
the two-body random ensemble (TBRE) \cite{Johnson1}.
In the studies of the 0 g.s. dominance we found 
a new type of cluster  which is given by the pairing
interaction with the largest multiplicity.

\section{THE $\alpha$ CLUSTERING PICTURE BASED ON THE SHELL MODEL }

The $^4$He ($\alpha$ cluster) has the $(0s)^4$ configuration in the zero-th order
approximation which belongs to the $\left[4 \right]$ symmetry of
SU(4).  As for the nucleus $^{20}$Ne, the configuration 
space for valence nucleons is given by $(1s0d)^4$. 
The $\left[4 \right]$ symmetry dominates 
in the wavefunctions obtained
by diagonalizing a shell model Hamiltonian with central
two-body interactions and one-body spin-orbit interaction,
as shown in Table 1 \cite{Arimax}. This indicates that there is a strong resemblance
between the shell model wavefunctions and the $\alpha$ cluster wavefunctions
of the nucleus $^{20}$Ne.

\begin{table}[htb]
\caption{  Percentage analysis of wavefuncations into different  orbital
symmetries.}
\label{table:4}
\begin{tabular}{cccccccc} \hline  
$T=0$ &   &    &    &    &   &     &  \\ \hline
$J$ &  $0_1$  & $0_2$   & $2_1$   & $2_2$   & $4_1$   & $4_2$    & $6_1$  \\ 
$\left[ 4 \right]$    & 91.8  & 92.0 & 89.9 & 81.9 & 87.5 & 83.0 & 80.0  \\
$\left[ 31 \right]$   & 7.9   & 7.3  &  9.7 & 16.5 & 11.8 & 15.7 & 18.8  \\
$\left[ 22 \right] $  &  0.2  &  0.6 &  0.3 &  1.4 &  0.5 &  0.7 &  1.3  \\
$\left[ 211 \right] $ &  0.1  &  0.1 &  0.1 &  0.2 &  0.2 &  0.6 &  0.4 \\
$\left[ 1111 \right] $&  &  & 0.0  & 0.0 &  &  &   \\
  \hline                                   
\end{tabular}
\end{table}

The matrix element of the Majorana operator $P_M$ between the ground state
of $^8$Be 
obtained by using the Cohen-Kurath interaction is $-5.77$.
If the SU(4) symmetry were exact, this value would be $-6$ for
the symmetry $\left[ 4 \right]$. The overlap between the ground
state obtained by the Cohen-Kurath interaction and that
by the Majorana interaction with the SU(4) symmetry $\left[ 4 \right]$
is 0.97.  Therefore, the Cohen-Kurath interaction favors
the SU(4) symmetry.

The Elliott SU(3) wavefunctions of the nucleus $^8$Be are
known to be identical to
the Wildermuth $\alpha$-cluster wavefunctions 
\begin{equation}                                       
 \Psi ( ^8{\rm Be}, (0s)^4 (0p)^4 [ 4 4 ] (40) IM ) =
{\cal N}   {\cal A} \left[  \Phi (\alpha_1) \Phi (\alpha_2) R_{nl} (r_{12})
 {\rm Y}_M^I (\theta_{12} \phi_{12})
          \right]   ~,
\end{equation}
where $R_{nl}(r_{12})$ are harmonic oscillator wavefunctions with
$2n+l=4$, and ${\cal N}$ is a normalization factor.
If we require $R_{nl}$ to satisfy the following
equation
\begin{equation}
 H {\cal A}
            \left[ \Phi (\alpha_1) \Phi (\alpha_2) R_{nl} (r_{12})
 {\rm Y}_M^I (\theta_{12} \phi_{12})
            \right]
=  E {\cal A}
            \left[ \Phi (\alpha_1) \Phi (\alpha_2) R_{nl} (r_{12})
 {\rm Y}_M^I (\theta_{12} \phi_{12})
            \right] ~, 
\end{equation}
$  {\cal A}
            \left[ \Phi (\alpha_1) \Phi (\alpha_2) R_{nl} (r_{12})
 {\rm Y}_M^I (\theta_{12} \phi_{12})
            \right]$  becomes the Resonating Group wavefunctions.
We can  say 
that the SU(3) shell model wavefunction provides
a  good approximation to that of the 
Generator  Coordinate Method (=Resonating Group method). 
We can see this from the following scenario. For $I=0$ 
\begin{eqnarray}
 {\rm e}^{ - (\alpha + \epsilon ) r_{12}^2 } &=& ( 1 -
 \epsilon r_{12} + \frac{\epsilon^2}{2} r_{12}^2 -
 \frac{\epsilon^3}{6} r_{12}^3 + \cdots ) \nonumber \\
 & \sim & R_{0s} + \epsilon R_{1s} + \epsilon^2 R_{2s} + \epsilon^3
R_{3s} + \cdots ~, 
\end{eqnarray}
one thus obtains 
\begin{equation}
  {\cal A}
            \left[ \Phi (\alpha_1) \Phi (\alpha_2) )
 {\rm e}^{ - (\alpha + \epsilon ) r_{12}^2 }
            \right]
=   {\cal A}
            \left[ \Phi (\alpha_1) \Phi (\alpha_2)
( R_{0s} + \epsilon R_{1s} + \epsilon^2 R_{2s} + \epsilon^3
R_{3s} + \cdots ) 
            \right] ~. \label{eq4}
\end{equation}
Because the Pauli principle forbids $R_{02}$ and $R_{1s}$ components, 
Eq. (\ref{eq4}) is reduced to 
\begin{equation}
  {\cal A}
            \left[ \Phi (\alpha_1) \Phi (\alpha_2) )
 {\rm e}^{ - (\alpha + \epsilon ) r_{12}^2 }
            \right]
\sim \epsilon^2  {\cal A}
            \left[ \Phi (\alpha_1) \Phi (\alpha_2)
(   R_{2s} + \epsilon R_{3s} + \cdots ) 
            \right] ~. 
\end{equation}
Absorbing  $\epsilon$    into the 
normalization factor, Eq. (5) is reduced to 

\begin{equation}
  {\cal A}
            \left[ \Phi (\alpha_1) \Phi (\alpha_2) )
R(r_{12}) 
            \right]
\sim   {\cal A} 
            \left[ \Phi (\alpha_1) \Phi (\alpha_2)
  R_{2s}    \right] ~,  \label{eq6}
\end{equation}
We start from
a 2-$\alpha$ condensate wavefunction
${\cal N} \left( {\cal A}
    \left[ \Phi (\alpha_1) \Phi (\alpha_2) ) {\rm e}^{- \beta r_{12}^2 }
    \right]
          \right)$ where $\beta = \alpha+ \epsilon$, and  have 
\begin{equation}
{\cal N}   {\cal A}
    \left[ \Phi (\alpha_1) {\rm e}^{- \beta r_{1}^2 } \Phi (\alpha_2) ) {\rm e}^{- \beta r_{2}^2 }
    \right]
= {\cal N}   {\cal A}
    \left[ \Phi (\alpha_1) \Phi (\alpha_2) ) {\rm e}^{- \beta r_{12}^2 }
    \Psi_{\rm C.M.} (\frac{\vec{r}_1 + \vec{r}_2}{2})
    \right]  ~, \label{eq7}
\end{equation}
where $\Psi_{\rm C.M.} (\frac{\vec{r}_1 + \vec{r}_2}{2})$ is center-of-mass
wavefunction of 2 $\alpha$ in a single spherical orbit $0S$.

As shown in Eq. (\ref{eq6}), the right hand side of
Eq. (\ref{eq7}) can be approximated by the
SU(3) shell model wavefunction when $\epsilon$ is small. 
Thus we have proved that the 2$\alpha$-condensate wavefunctions are the same
as those of the shell model if $\epsilon$ is small.
However, when $\epsilon$ is not very small, higher configurations of the 
shell model space must contribute to modify the shell model
wavefunctions. In such  cases the $\alpha$-cluster  model is more efficient
than the shell model. One
may similarly discuss the 3-$\alpha$ condensate states.

\section{Towards understanding of the 0 g.s. dominance}

The ground state spins$^{\rm parity}$ of even-even nuclei are always 0$^+$. 
We believed that this fact   originates from
  short range attractive interaction. 
However, the   (0 g.s.)   was discovered by using
the two-body random ensemble (TBRE)   \cite{Johnson1}.
 There have been many  efforts to understand
this interesting and important observation
\cite{Bijker0,Zhao2}.

Here we exemplify our work  \cite{Zhao2} 
by the case of four fermions in a single-$j$ shell.
The Hamiltonian  for fermions in a
single-$j$ shell is defined as follows
\begin{eqnarray}
&& H = \sum_J G_J  A^{J \dagger} \cdot A^{J} \equiv  
\sum_J \sqrt{2J+1} G_J 
\left( A^{J \dagger} \times
\tilde{A}^J \right)^0, \nonumber \\
&&  A^{J \dagger} = \frac{1}{\sqrt{2}} \left( a_j^{\dagger} \times a_j^{\dagger}
 \right)^J, ~ ~
 \tilde{A}^J = - \frac{1}{\sqrt{2}} \left( \tilde{a}_j \times
 \tilde{a}_j \right)^J, ~ ~ G_J = \langle j^2 J|V| j^2 J \rangle.  \label{pair}
\end{eqnarray}
$G_J$'s are taken as a
set of Gaussian-type random numbers with  
a width being 1 and an average being 0. This two-body random ensemble
is referred to as ``TBRE". 
The $I$ g.s. probabilities in this paper are obtained by 1000 runs
of a TBRE Hamiltonian.

In Ref. \cite{Zhao2} we introduced an empirical formula to predict
the $P(I)$'s, which are the probabilities of a state with spin $I$ to be the
ground state. This formula was found to be valid for  both a 
single-$j$ shell and many-$j$ shells, for both an even value 
of particle number and an odd value of particle number, and for both 
fermions and bosons.

Our procedure to predict the $P(I)$'s is as follows.
Let us set only one of the $G_J$'s equal to $-1$ and the others to zero,
and find the spin $I$ of the ground state. We 
repeat this process for 
all two-body interactions $G_J$. 
We can find how many times the ground state has angular momentum $I$. 
This number is denoted as $N_I$. The values of $N_I$ 
for four nucleons in a single-$j$ shell can be
easily counted  from  Table 2 when $j$ is not very large.

\begin{table}[htb]
\caption{  The angular momenta $I$'s which give the lowest eigenvalues
for 4 fermions in a single-$j$ shell, 
when  $G_J=-1$  and all other parameters are  0.    }
\label{table:6}
\begin{tiny}
\begin{tabular}{ccccccccccccccccc} \hline  
$2j$ &  $G_0$ &  $G_2$ &  $G_4$ &  $G_6$ &  $G_8$ &  $G_{10}$ &  $G_{12}$
&  $G_{14}$ &  $G_{16}$ &  $G_{18}$ &  $G_{20}$ &  $G_{22}$ &  $G_{24}$
&  $G_{26}$ &  $G_{28}$ &  $G_{30}$   \\  \hline
7  & 0 &4 &2 &8 &   &   & & & &  & & & & & & \\
9  & 0 &4 &0 &0 &12 &   & & & &  & & & & & & \\
11 & 0 &4 &0 &4 &8  &16 & & & &  & & & & & & \\
13 & 0 &4 &0 &2 &2  &12 &20 & & &  & & & & & & \\
15 & 0 &4 &0 &2 &0  &0  &16 &24 & &  & & & & & & \\
17 & 0 &4 &6 &0 &4  &2  &0  &20 &28 &  & & & & & & \\
19 & 0 &4 &8 &0 &2  &8  &2  &16 &24 &32 & & & & & & \\
21 & 0 &4 &8 &0 &2  &0  &0  &0  &20 &28 &36 & & & & & \\
23 & 0 &4 &8 &0 &2  &0  &10 &2  &0  &24 &32 &40 & & & & \\
25 & 0 &4 &8 &0 &2  &4  &8  &10 &6  &0  &28 &36 &44 & & & \\
27 & 0 &4 &8 &0 &2  &4  &2  &0  &0  &4  &20 &32 &40 &48 & & \\
29 & 0 &4 &8 &0 &0  &2  &6  &8  &12 &8  &0  &24 &36 &44 &52 & \\
31 & 0 &4 &8 &0 &0  &2  &0  &8  &14 &16 &6  &0  &32 &40 &48 & 56 \\
  \hline 
\end{tabular}
\end{tiny}
\end{table}

Using the $N_I$, we can predict the probability that the ground state 
has angular momentum $I$ as $P(I) = N_I/N$, where $N$ is the number of 
independent two-body matrix elements $(N=j+1/2$ for fermions in a 
single-$j$ shell). A nice agreement between our predicted $P(0)$'s and
those obtained by diagonalizing a TBRE Hamiltonian is shown
in Fig. 1. Comparison for more complicated cases can be found
in \cite{Zhao2}.

For the case of fermions in a single-$j$ shell, one easily notices
that the $P(I_{\rm max})$ is sizable. The reason is very simple:
$N_{I_{\rm max}} \equiv 1$, which comes from  the $G_{J_{\rm max}}$
term. Therefore, we predict that  the $P(I_{\rm max})= 1/N$, where
$N=j+\frac{1}{2}$ for a single-$j$ shell. Fig. 2 shows
a comparison between the predicted $P(I_{\rm max})$'s by $1/N$ and
those by the TBRE Hamiltonians. We see a remarkable
agreement.

The above empirical formula also provides us with 
other very important insights. 
It presents a guideline to tell  
which interactions are essential to produce a sizable  
$P(I)$ in a many-body system.
For   example,   $P(0)$ for $j=31/2$ is given
essentially by the two-body matrix elements with
 $J=0$, 6, 8, 12, and 22. The $P(0)$ 
would be close to zero without these five terms.  This disproves a
popular idea that the angular momentum 0 ground state  (0 g.s.)  
dominance may be independent of two-body interactions.

The 0 g.s. dominance has not been   understood 
yet from a more sophisticated level. Further works 
are therefore warranted.

\section{A NEW TYPE OF CLUSTER}

When one examines the eigenvalues and wavefunctions
for $G_{J_{\rm max}} =-1$ and other $G_{J'}=0$ for
a Hamiltonian of Eq. (\ref{pair}), one 
finds a very interesting phenomenon: the states
can be classified by clusters. Here we discuss systems
 up to four particles ($n=4$) with $j=31/2$. 
 Systems with more particles exhibit a similar
 behavior. 

For $n=2$, there are only two cases: $I=30$
which has $E=-1$ and $I\neq 30$ which have
$E=0$.

For $n=3$, it was proved \cite{preprint} 
that the nonzero eigenvalue of each $I$ states
for $G_{J_{\rm max}} =-1$ and other $G_{J'}=0$
is given by the configuration 
 $ |\Psi^{E\neq 0}_{n=3} \rangle = \frac{1}{\sqrt{N^I_{jJ_{\rm max}}} }
$$ \left( a_j^{\dagger} \times A^{J=J_{\rm max}} \right)^I |0 \rangle$,
 where $N^I_{jJ_{\rm max}}$ is the normalization factor.
Namely, for each $I$ there is one and only one
state which has non-zero eigenvalue. One can prove that 
this value is very close to $-1$ unless $I\sim I^{(3)}_{\rm max}$
(The $I^{(n)}_{\rm max}$  refers to the maximum angular momentum
of $n$ particles in a single-$j$ shell). Therefore, the first 
particle $a_j^{\dagger}$ in $|\Psi^{E\neq 0}_{n=3} \rangle$
behaves like a ``spectator", and
the $A^{J_{\rm max} \dagger }$ behaves like a two-particle ``cluster".
Eigenvalues for those $J \neq J_{\rm max} = 2j-1$ in
$|\Psi^{E}_{n=3} \rangle$ are zero.

Now let us come to the case of $n=4$. 
In  \cite{Gino} it was proved
that  the eigenvalues $E$ of $n=4$ are asymptotically
0, $-1$ or $-2$ for small $I$,  
and that the  states with $E\sim 0$, $-1$ or $-2$ are constructed by 
zero, one or two pairs with spin $J=J_{\rm max}$.

Besides these ``integer" eigenvalues, ``non-zero" eigenvalues
arise as  $I$ is larger than $2j-9$. The values of these ``non-integer" 
eigenvalues are very close to those of $n=3$ states with 
$I^{(3)} \sim I^{(3)}_{\rm max}$. The configurations for these 
states can be approximately given by one cluster consisting of three 
particles with $I^{(3)} \sim I^{(3)}_{\rm max}$
and one spectator. We shall explain this in more details below.

Fig. 3(a)-(c) plots  the distribution  of all 
non-zero eigenvalues for $n=3$ and 4.
From Fig. 3, one sees that 
these eigenvalues are  converged
at  a few values but with exceptions. The ``converged" 
values are {\it very} close to those of eigenvalues 
of $n=3$.

\begin{table}[htb]
\caption{ The lowest  eigenvalue ${\cal E}_I$
(columns ``SM" and ``coupled") for 
$I\ge 28 $ states of 
four fermions in a $j=31/2$ shell 
and its overlap (last column) between the wavefunction obtained by 
the shell model calculation and that obtained by  
coupling  $a_j^{\dag}$ to the $I^{(3)}_{\rm max}$ state. 
The column ``(SM)" is obtained by the shell model 
diagonalization, and the    ``$F_I$" is the matrix element 
of $H_{J_{\rm max}}$ in the configuration of 
coupling  $a_j^{\dag}$ to the $I^{(3)}_{\rm max}$ state. 
Italic font is used for the overlaps  which are
 not close to 1. }

\vspace{0.2in}

\begin{tabular}{c|ccc} \hline \hline
 $I$ & ${\cal E}_I$ (SM) ~  & $F_I$ (coupled) & overlap  \\ \hline
28 & -2.26271186440690  & -2.262711864406782 & 1.000000000000000 \\
29 & -2.26271186440682  & -2.262711864406777 & 1.000000000000000 \\
$\vdots$ & $\vdots$  & $\vdots$ & $\vdots$    \\
43 & -2.26272031287460  & -2.262720286322401 & 0.999999987392690    \\
44 & -2.26317530567842  & -2.262776481261782 & 0.999151747579904    \\
45 & -2.26282037299297  & -2.262819747102017 & 0.999999632523561    \\
46 & -2.26963309159052  & -2.263514816015588 & 0.982828211942919    \\
47 & -2.26378385186917  & -2.263772302947436 & 0.999992036003522    \\
48 & -2.34719850307215  & -2.270625142453812 & {\it 0.780582505446094}    \\
49 & -2.27068252318197  & -2.270571840272616 & 0.999929221753443    \\
50 & -2.57872583562800  & -2.323429204525185 & {\it 0.706859839896674}    \\
51 & -2.30488200470359  & -2.304882004703592 & 1.000000000000000 \\
52 & -2.89017281282010  & -2.592166600952603 & {\it 0.873170713095796}    \\
53 & -2.41926851025870  & -2.419268510258698 & 1.000000000000000 \\
54 & -3.24511394047522  & -3.245113940475225 & 1.000000000000000 \\
56 & -3.66369313113292  & -3.663693131132918 & 1.000000000000000 \\
 \hline  
\end{tabular}
\end{table}

For $j=31/2$ and $n=4$ the total number of states
is 790. The number of states with non-zero eigenvalues
is 380. Within a precision $10^{-2}$, 
308 states which eigenvalues  are located at the eigenvalues of $n=3$, 
and 21 states have eigenvalues at $-2$.
It is noted that most of exceptions of eigenvalues
of $n=4$ can be nicely given by
a three-particle cluster coupled with a single-$j$ particle
in a high precision.
In this example only
four states with $I=48$, two states with $I=46$, two states
with $I=44$ are not given well by
the  pictures of
a three-particle cluster coupled with a single-$j$ particle
or two pairs with spin $J_{\rm max}$.  These states behave as four-body 
clusters in systems with $n>4$.

Let us exemplify this  by the peak near $2.25$. 
$E_{I_{\rm max}}^{(n=3)}$=$-\frac{267}{118}$ =2.26271186440678
for $j=31/2$. 
The smallest $I$ which can be coupled by 
three particles with $I=I^{(n=3)}_{\rm max}$ and a single-$j$
spectator is   $I^{(3)}_{\rm max} -j=2j-3$ and here 28. 
The lowest eigenvalue ${\cal E}_I$ of $I=28$
obtained by a shell model diagonalization for $n=4$ 
is $-2.26271186440689$, which is very close to
$E_{I_{\rm max}}^{(n=3)}$. Some of the  ${\cal E}_I$ 
with $I$ between  $28$ to 56 are listed 
in Table 3.

We have calculated all overlaps between states 
of $n=4$ with energies focused in the peaks 
and those of simple configurations
obtained by coupling a non-zero energy cluster with 
$I^{(3)} \sim I^{(3)}_{\rm max}$ of three fermions and a single-$j$
particles, which shows similar situation as Table 1.
Therefore, we conclude that those     ``non-integer"
 eigenvalues of $n=4$ with
$H = H_{J_{\rm max}}$ are given in a high precision by  
three-particle cluster (nonzero energy) coupled with a single-$j$ particle.

\section{SUMMARY}

I first discussed  the $\alpha$-cluster picture for a few
typical nuclei from the view point  of the shell model.
Then I discussed an empirical approach of
predicting  the $I$ g.s. probabilities in the
presence of random interactions. 
A new type of clustering phenomenon was  discussed. 
We are now trying to find applications of this clustering phenomenon. 

I thank Drs. Y. M.  Zhao and N. Yoshinaga   for their collaborations  
in my work.

\newpage

Figure 1 ~~ {A comparison between our predicted $P(0)$'s and those
obtained by 1000 runs of a TBRE Hamiltonian. }

\vspace{0.3in}

Figure 2 ~~ {A comparison between the predicted $P(I_{\rm max})$
($\sim 1/(j+\frac{1}{2}$)  and those 
obtained by 1000 runs of a TBRE Hamiltonian. }

\vspace{0.3in }

Figure 3 ~~  
Distribution  of all non-zero eigenvalues for $n=4$. 
It is seen that these eigenvalues of $n=4$ are  converged 
at eigenvalues of $n=3$ but with very few exceptions.


\newpage


\begin{thebibliography}{19}
\bibitem{Arimax} T. Inoue, T. Sebe, H. Hagiwara, and A. Arima,
Nucl. Phys. {\bf 59} (1964) 1.

\bibitem{Johnson1}  C.W. Johnson, G.F. Bertsch and D.J. Dean 
     Phys. Rev. Lett. {bf 80} (1998) 2749. 

\bibitem{Bijker0}  
R. Bijker {\it et al.}, Phys.  Rev. {\bf C60} (1999) 021302;  
D. Mulhall {\it et al.}, Phys. Rev. Lett. {\bf 85}(2000)  4016;
D. Kusnezov, Phys. Rev. Lett.  {\bf 85} (2000)  3773. 
 R. Bijker {\it et al.}, Phys. Rev. {\bf C64} (2001)  (R)061303; 
 P. H-T. Chau {\it et al.}, Phys. Rev. {\bf C66}   (2002)  061302. 


\bibitem{Zhao2} Y.M. Zhao, A. Arima, and N. Yoshinaga, Phys. Rev. {\bf C66} (2002)  034302; 
ibid. {\bf C66} (2002)  064322; ibid. {\bf C68} (2003)  014322; 
nucl-th/0311050.


\bibitem{preprint} Y. M. Zhao and A. Arima, preprint  (unpublished). 

\bibitem{Gino} Y. M. Zhao and  A. Arima {\it et al.}, 
Phys. Rev. {\bf C68} (2003) 034320.



\end{thebibliography}
\end{document}